\documentstyle[psfig]{mn}
\def\H0{{\it H}$_0$}

\def\q0{{\it q}$_0$}
\def\kmps{km~s$^{-1}$}

\def\ergps{erg~s$^{-1}$}

\def\nH{$N_{\rm H}$} 
\def\psqcm{cm$^{-2}$}
\def\ergpspsqcm{erg~cm$^{-2}$~s$^{-1}$}

\def\ltsima{$\; \buildrel < \over \sim \;$}
\def\simlt{\lower.5ex\hbox{\ltsima}}
\def\gtsima{$\; \buildrel > \over \sim \;$}
\def\simgt{\lower.5ex\hbox{\gtsima}}

\def\AX{AX\thinspace J1749+684\ }
\def\AXc{AX\thinspace J1749+684}

\def\RX{RX\thinspace J\thinspace 174949+682303\ }
\def\RXc{RX\thinspace J\thinspace 174949+682303}

\title[Narrow emission-line galaxy \AXc]
{\AXc: a narrow emission-line galaxy with a flat X-ray spectrum}
\author[K. Iwasawa et al]
{\parbox[]{6.5in}{K. Iwasawa,$^1$ A.C. Fabian,$^1$ W.N. Brandt,$^2$
 \thanks{Present address: The Pennsylvania State University, Department of Astronomy and Astrophysics,
525 Davey Park, PA 16802, USA
}
C.S. Crawford,$^1$ and O. Almaini$^1$}\\
\\
$^1$ Institute of Astronomy, Madingley Road, Cambridge CB3 0HA\\
$^2$ Harvard-Smithsonian Center for Astrophysics, 60 Garden Street, Cambridge MA 02138, USA\\} 

\date{}

\begin{document}
\date{MNRAS in press}
\maketitle

\begin{abstract}

\noindent We report the serendipitous detection of an X-ray source, \AXc\, with
the ASCA Gas Imaging Spectrometer.  \AX is identified with a
LINER/starburst-type spiral galaxy KUG\thinspace 1750+683A at a
redshift $z = 0.05$.  It has a hard X-ray spectrum, consistent with
that of the X-ray background (XRB) in the 1-10~keV band.  Despite the
optical classification, the X-ray luminosity cannot be explained by
starburst activity. Combined with spatial variations in the optical
emission line ratios, this suggests the presence of an obscured
Seyfert nucleus embedded within a starforming galaxy.  Similar
behaviour could explain the ambiguous properties of the faint
narrow-line X-ray galaxies (NLXGs) emerging from deep X-ray surveys.

\end{abstract}

\begin{keywords}
galaxies: individual: \AX --
galaxies: active -- 
X-rays: \end{keywords}

\section{INTRODUCTION}

Although the origin of much of the soft XRB below $\sim 2$~keV is now
understood in terms of the integrated contribution of quasars and
faint NLXGs (Boyle et al 1995; Roche et al 1995; Carballo et al 1995;
Griffiths et al 1996; McHardy et al 1996; Hasinger 1997), the
situation is less clear at higher energies where the bulk of the
energy density occurs. A strong possibility is that the NLXGs have
hard X-ray spectra similar to that of the XRB, which is well fitted by
a power-law of photon index $\Gamma=1.4$ in the 1-7~keV band (Gendreau
et al 1995; Chen et al 1997).  The NLXGs may therefore dominate the
background contribution at harder energies.  In the softer
0.1--2.4~keV ROSAT band there are already indications that NLXGs show
significantly harder spectra than other types of X-ray sources
(Almaini et al 1996, Romero-Colmenero et al 1996).  The hard spectra
of NLXGs may be intrinsic to the continua of those sources, making
them unlike any other well-studied objects, or it may be produced by
the integrated effect of varied levels of intrinsic absorption in a
more typical active galaxy population. The 2--10 keV spectra of
sources detected serendipitously are therefore of great interest since
they may be the brighter, and probably nearer, members of the
population dominating the XRB. Studies of such objects may therefore
help to reveal the true nature of this X-ray population.

So far most NLXGs have been discovered in deep ROSAT observations,
where they are now thought to dominate the X-ray source population at
faint fluxes below
$S(0.5-2.0$\,keV$)=4\times10^{-15}$erg$\,$s$^{-1}$cm$^{-2}$.  A small
number of bright NLXGs have been known for many years (see
e.g. Piccinotti et al 1982; Lawrence \& Elvis 1982) but their
relevance to the XRB is unclear. Serendipitous sources detected with
ASCA provide a new way forward, since good spectra are then obtained
over the whole 1-10~keV band. The discovery of possible flat X-ray
spectrum from NGC 3628 (Yaqoob et al 1995) has already highlighted the
possible importance of previously hidden AGN in explaining the hard
XRB.

We report here on the serendipitous detection of a previously
unidentified X-ray source with a flat X-ray spectrum, discovered
during the ASCA observation of Mrk~507.  The object is $\sim 20$
arcmin away from Mrk~507 and seen in the same Gas Imaging Spectrometer
(GIS) field of view. We refer to this source as \AXc. A ROSAT Position
Sensitive Proportional Counter (PSPC) image of the same field greatly
improved the positional uncertainty of the source and allowed us to
identify it with a galaxy which has a UV excess.  An optical spectrum
of \AX then allows us to obtain the redshift and deduce the nature of
the activity in this galaxy. Our results show that it is a low
redshift and hence one of the brightest ($\sim 10^{-12}$erg
cm$^{-2}$s$^{-1}$) members of the NLXG population.

A value of the Hubble parameter of $H_0=50$ km s$^{-1}$ Mpc$^{-1}$ and
a cosmological deceleration parameter of $q_0={1\over 2}$ have been
assumed throughout.

\section{X-ray data}

\subsection{Observations}

ASCA observed Mrk~507 (and thus \AXc) on 1995 December 16 with a total
exposure time of 34.8 ks with the GIS. The data reduction and analysis
were performed using {\sc ftools} and {\sc xspec}, provided by the
ASCA Guest Observer Facility at Goddard Space Flight Center. Spectral
data were taken from circular regions with 5 arcmin radii, giving a G2
count rate of $9.9\times10^{-3}$ counts s$^{-1}$. Only the GIS data
are available for this source, since it was outside the SIS field of
view. This object was also observed in a ROSAT PSPC pointing at
Mrk~507 on 1993 August 8--10. The total exposure time for this
pointing was 24.7 ks.  The brightest source in this ASCA observation
is Kaz 163, a Seyfert 1 galaxy at a redshift of $z=0.063$, $\sim 10$
arcmin SW of Mrk~507.  This object will be used for comparison in the
spectral fitting.  For the results of spectral fits, we quote 90 per
cent confidence errors for one parameter of interest.

\subsection{Source identification}

\AX is detected serendipitously about 21 arcmin south of Mrk~507 with
the GIS. The source position determined from a summed G2+G3 image is
$\alpha_{2000} = 17^{\rm h} 49^{\rm m} 46^{\rm s}$, $\delta_{2000} =
+68^{\circ} 23^{\prime} 46^{\prime\prime}$.  We estimated that the
positional uncertainty is probably $\sim$1--2 arcmin (see Fig. 1; also
Gotthelf 1996).

The ROSAT PSPC has much better positional accuracy than ASCA, and we
have searched the PSPC field in the 0.5--2.0 keV band for a
counterpart. One PSPC source within the ASCA error box has a similar
X-ray spectrum and a comparable flux level. It is the brightest ROSAT
source in the ASCA error circle. The position of the ROSAT source is
$\alpha_{2000} = 17^{\rm h} 49^{\rm m} 49^{\rm s}$, $\delta_{2000} =
+68^{\circ} 23^{\prime} 03^{\prime\prime}$ with an error circle radius
of about 20 arcsec, and we shall refer to this source as \RXc.

Several objects are seen around \AX on the Palomar Optical Sky Survey
(POSS-I) plate (Fig. 1). A spiral galaxy, KUG~1750+683A ($V = 16$
mag), which was selected by the ultraviolet excess galaxy survey with
the Kiso Schmidt camera (Takase \& Miyauchi-Isobe 1989), is the
closest object to the ROSAT position for \RX and the only one within
its error circle. We regard this galaxy to be the most probable
counterpart to both \AX and \RXc.  We note that a ROSAT source is also
associated with the galaxy KUG~1750+683B (see Figure 1), but with 10
times less ROSAT flux than KUG~1750+683A we do not consider this to be
a likely counterpart to \AX.

\subsection{X-ray spectrum}

We can use the X-ray spectrum of the nearby Seyfert 1 galaxy Kaz~163
to compare with \AX.  Further details of the ASCA spectrum of Kaz~163
are given in Iwasawa et al (1997).  The object has a featureless X-ray
spectrum with a spectral slope typical of Seyfert 1 galaxies
($\Gamma=2$). The spectral ratio between \AX and Kaz~163 from the GIS
data shows that \AX has a considerably harder X-ray spectrum (see
Fig. 2). Results of a power-law fit to the two GIS spectra are shown
in Table 1. Detection of excess absorption in the ASCA data alone is
not statistically significant. The ROSAT PSPC spectrum is complicated
by the presence of a detector support ring next to the image of the
source. The source is very faint in the ROSAT soft-band image
($<0.4$~keV) however, suggesting the presence of an intrinsic
absorbing column. The ASCA GIS spectrum gives $N_{\rm H}=8\pm3\times
10^{21}$\psqcm\, for $\Gamma=2$.  If we assume that the source does
not vary and make a joint fit of the ROSAT and ASCA data we do obtain
a statistically acceptable result (reduced $\chi^2=0.98$) for $N_{\rm
H}=7.4\times 10^{21}$\psqcm\, and $\Gamma=1.63$. The joint
uncertainties on $N_{\rm H}$ and $\Gamma$ are shown in Fig.~3. We note
that a model dominated by X-ray reflection gives an equally good fit
to the joint spectrum, but still requires a comparable level of
neutral absorption. The GIS data does not place a useful constraint on
the equivalent width of an iron K line, giving a 90 per cent upper
limit of $\sim$ 1.7keV.

\section{Optical data}

\begin{table*}
\caption{Results of spectral fits to the ASCA data on Kaz\thinspace
163 and \AXc.  A power-law modified by cold absorption is fitted. The
values of absorption quoted here are excess column density above
Galactic value (\nH = $4.3\times 10^{20}$\psqcm).}
\begin{center}
\begin{tabular}{lcccccccc}
Object & $z$ & $\Gamma$ & \nH & $\chi^2$/dof & $f_{\rm 0.6-2keV}$ &
$f_{\rm 2-10keV}$ & $L_{\rm 2-10keV}$ & EW(FeK) \\
&&& $10^{21}$\psqcm\,  &&\multicolumn{2}{c}{$10^{-12}$\ergpspsqcm} &
$10^{43}$\ergps & keV \\[5pt]
\AX & 0.050 & $1.40^{+0.55}_{-0.37}$ & $2.1^{+6.2}_{-2.1}$ & 53.80/57 &
0.20 & 0.96 & 1.0 & $<1.7$ \\
&& $1.23^{+0.21}_{-0.27}$ & --- & 54.28/58 &&&& \\
Kaz\thinspace 163 & 0.063 & $2.04^{+0.12}_{-0.96}$ & $0.67^{+0.57}_{-0.52}$ & 210.1/227 &
1.0 & 1.6 & 2.8 & $<0.6$ \\[5pt]
\end{tabular}
\end{center}
\end{table*}

\begin{figure}
%\vspace{7cm}
\centerline{\psfig{figure=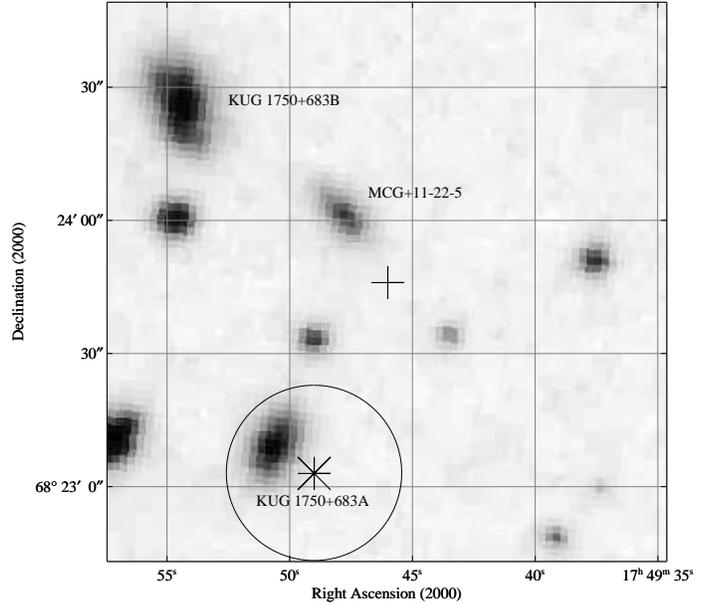,width=0.5\textwidth,angle=0}}
\caption{The Palomar Optical Sky Survey $B_{\rm J}$ plate around \AXc. 
The optical image is roughly the same size as the ASCA 
error region, and the cross shows the ASCA X-ray centroid. 
The ROSAT PSPC position for \RX (see the text) is 
marked with a star, and the circle shows the PSPC error 
circle. Galaxies listed in NED are labeled.}
\end{figure}

\begin{figure}
%\vspace{7cm}
\centerline{\psfig{figure=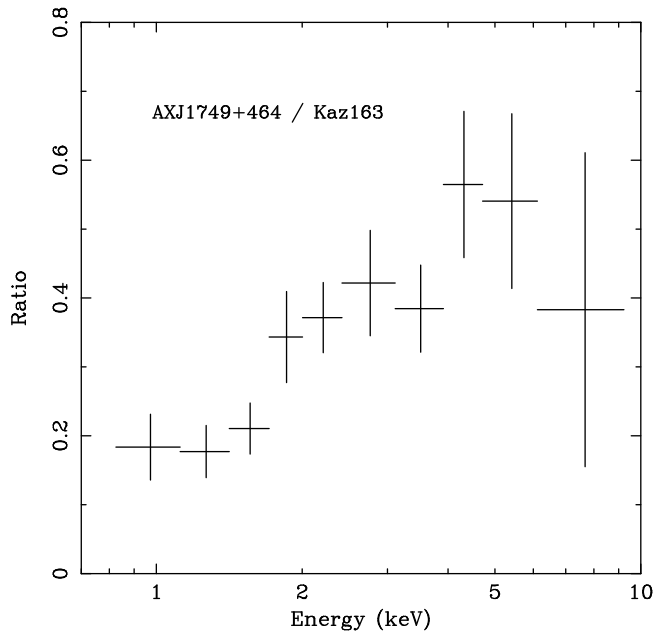,width=0.7\textwidth,angle=270}}
\caption{The spectral ratio of \AX and Kaz~163 from 
the ASCA GIS data. The X-ray spectrum of \AX is
significantly harder than typical Seyfert~1 galaxies like Kaz~163.}
\end{figure}

\begin{figure}
%\vspace{7cm}
\centerline{\psfig{figure=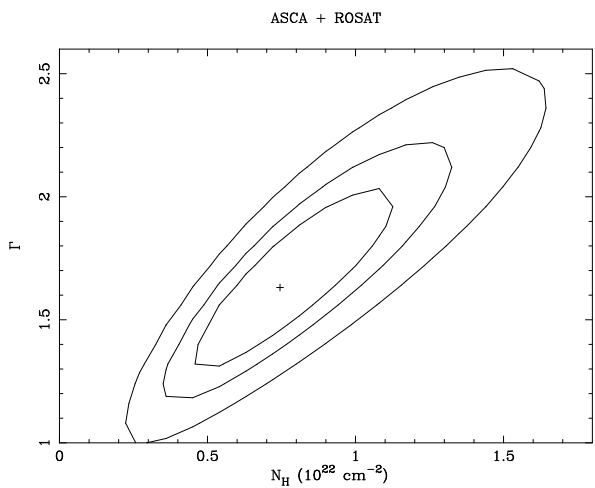,width=0.5\textwidth,angle=270}}
\caption{Confidence contours (corresponding to 68, 90 and 95
percent confidence levels) for $N_{\rm H}$ and $\Gamma$ obtained from a
joint fit of the ASCA and ROSAT data, assuming no variations in the
source flux over the 2 year interval between the observations.}
\end{figure}

\begin{figure}
\centerline{\psfig{figure=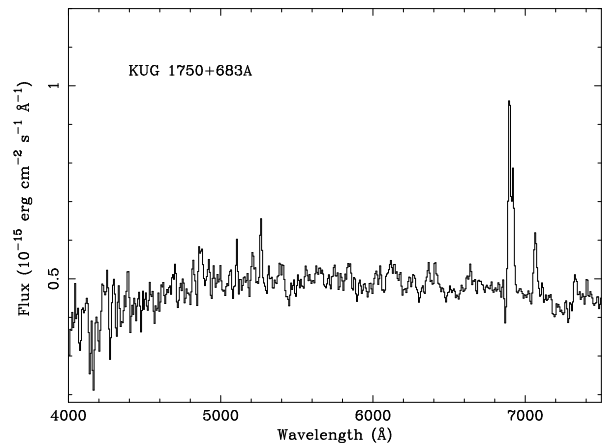,width=0.5\textwidth,angle=270}}
\caption{The optical spectrum of KUG1750+683A, the most likely optical
counterpart of \AXc. The spectrum is not corrected for the redshift
($z=0.05$) but is corrected for Galactic reddening $A_{\rm V}=0.30$. See
text for major diagnostic line ratios.}
\end{figure}

Following our X-ray detection, an optical spectrum of KUG\thinspace
1750+683A (Fig. 4) was obtained by one of us (ACF) at the Isaac Newton
Telescope using a slit of 1.35 arcsec width in 1.2 arcsec seeing.  A
1000s exposure was obtained in 1996 December  with the Intermediate
Dispersion Spectrograph, using the 235mm camera and R150V grating.  Major
emission lines are detected with no significant broadening (FWHM $<
550$\kmps), with the upper limit on the line widths arising because of
the instrumental resolution.  The redshift is determined to be $z =
0.050$.  Using the excitation diagrams of Veilleux \& Osterbrock
(1987), the emission line ratios [SII]$\lambda\lambda
6716,6731$/H$\alpha$ = 0.40, [NII]$\lambda 6583$/H$\alpha$ = 0.58 and
[OIII]$\lambda 5007$/H$\beta$ = 2.21, place this object in an
intermediate region between LINERs and HII-type objects.  Strong
[OI]$\lambda 6300$, characteristic of LINERs, is not detected
([OI]$\lambda6300$/H$\alpha < 0.2$). The integrated H$\alpha$ flux in
the slit is $8.86^{+0.28}_{-0.38}\times 10^{-15}$\ergpspsqcm.

A large Balmer decrement, H$\alpha$/H$\beta = 7.32$, suggests
relatively large reddening in the narrow emission-line region. This
corresponds to a column of $\sim 5\times 10^{21}$\psqcm\, if we assume
case B recombination for the intrinsic line ratio and the Galactic
dust-to-gas ratio (Bohlin, Savage \& Drake 1978). This column density
is comparable to that found from the joint ASCA/ROSAT spectral fitting
(Fig.~3).

KUG\thinspace 1750+683A thus appears to be a typical NLXG,
distinguished only by its proximity and brightness. We note however
that several of the emission lines are extended and the line ratios
vary with position (Fig.~5). We have therefore extracted the important
line ratios [NII]$\lambda 6583$/H$\alpha$ and [SII]$\lambda\lambda
6716,6731$/H$\alpha$ as functions of position along the slit (Fig.~6).
These ratios peak around cross-section 130, 1.8 arcsec (2.4 kpc) from
the continuum peak of the galaxy, at values greater than those for the
integrated spectrum.  The line ratio behaviour is similar to that seen
in the Seyfert 1 galaxy NGC~3227 (Delgado \& Perez 1997) and is
suggestive of photoionization by an active nucleus.  It is therefore
plausible that KUG\thinspace 1750+683A is a moderately obscured AGN
surrounded by an active starforming galaxy, giving intermediate line
diagnostics when an integrated spectrum is considered.

We note that KUG 1750+683A is also a radio source (VLA 1749.8+6823)
detected in the 1.5 GHz VLA NEP survey (Kollgaard et~al.  1994) with a
radio flux density 7.7 mJy. Such a radio flux could arise from either
the AGN or starburst component.

\begin{figure}
\centerline{\psfig{figure=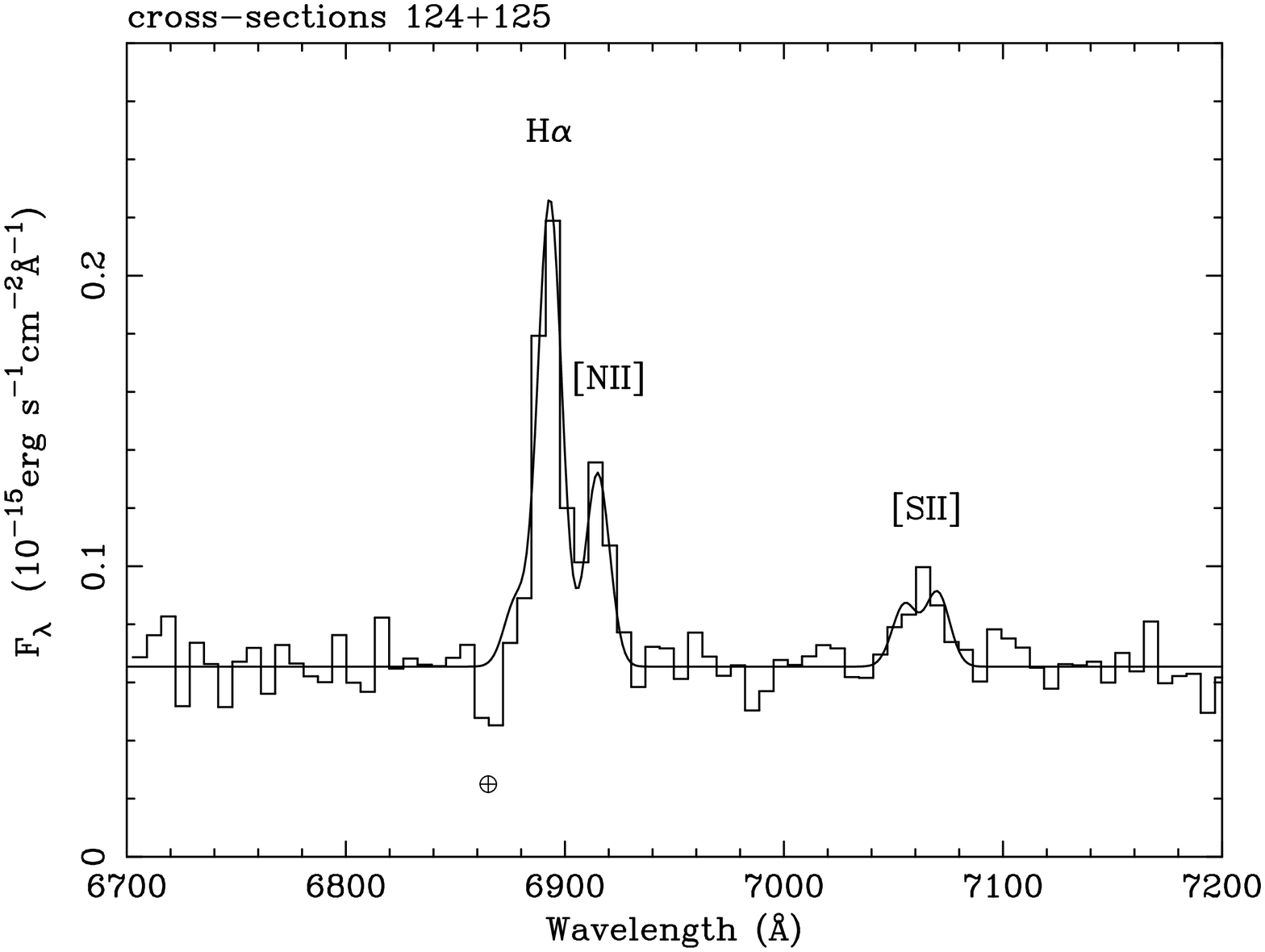,width=0.5\textwidth,angle=270}}
\centerline{\psfig{figure=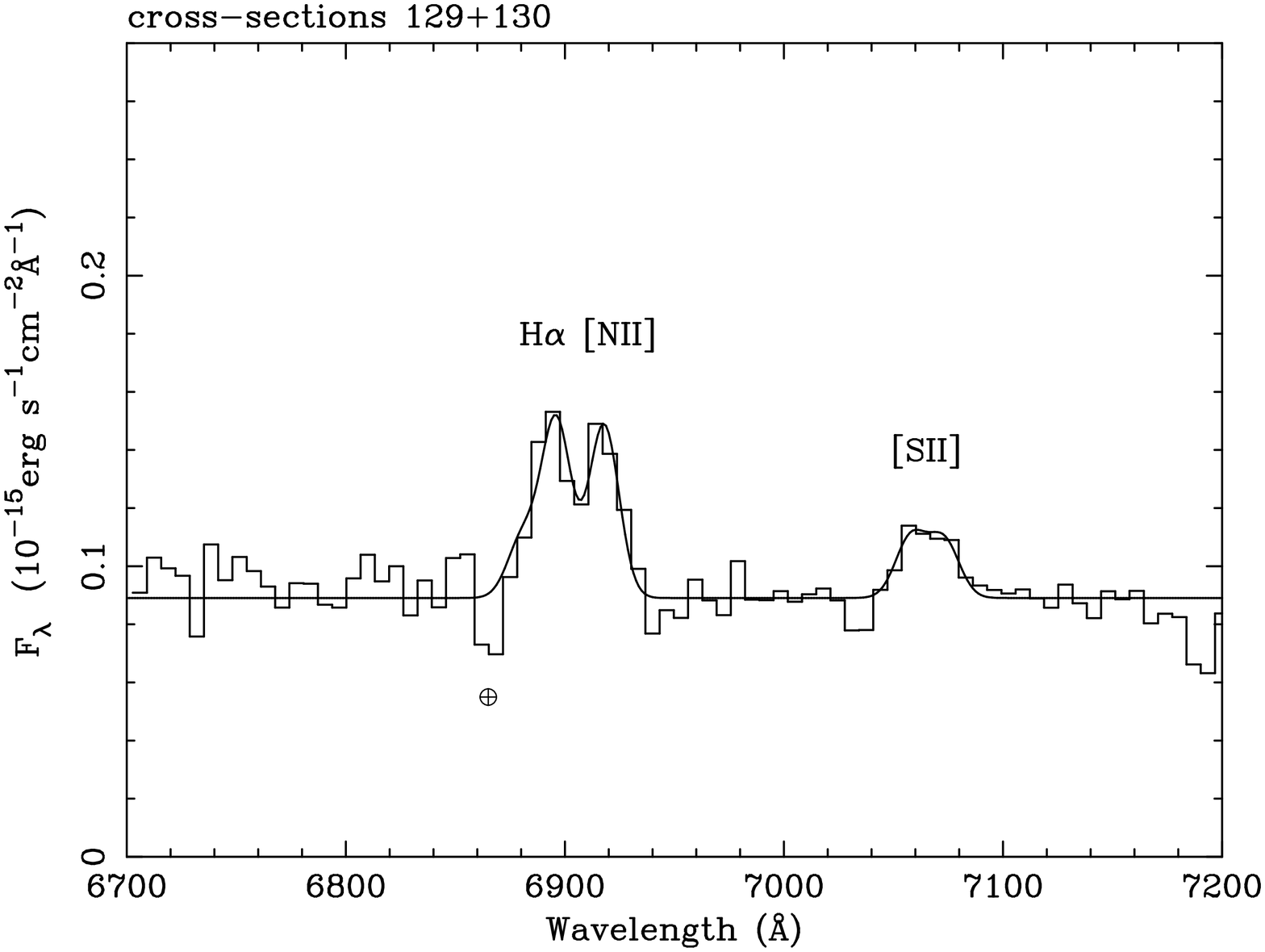,width=0.5\textwidth,angle=270}}
\caption{Spectrum around [NII]$\lambda 6583$/H$\alpha$ and
[SII]$\lambda\lambda 6716,6731$/H$\alpha$ from cross-sections 124+125
(top) and 129+130 (below).  These spectra are taken from regions
spatially separated by 3.5 arcsec (4.7 kpc).  Note the large change in
the strength of H$\alpha$ with respect to the other lines.  The
absorption feature at 6870\AA\, is atmospheric.  }
\end{figure}

\begin{figure}
\centerline{\psfig{figure=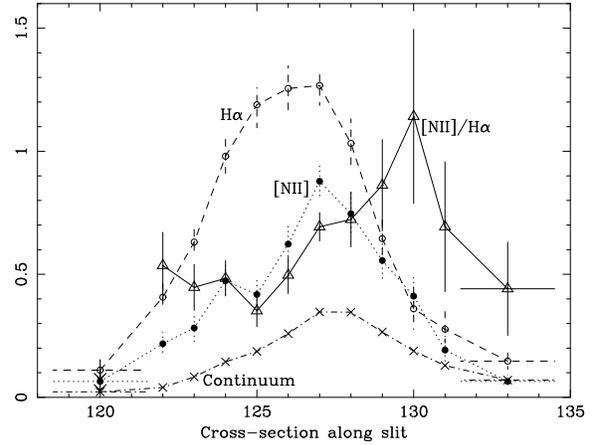,width=0.5\textwidth,angle=270}}
\caption{The spatial intensity profile of H$\alpha$, [NII]$\lambda
6583$, the neighbouring continuum and [NII]/H$\alpha$ along the slit
(each cross-section is 0.7 arcsec, or 0.94 kpc).  The y-axis gives the
line ratio for [NII]/H$\alpha$. The line fluxes are given in units of
$1\times10^{-15}$erg$\,$s$^{-1}$cm$^{-2}$ while the continuum is given
in units of $5\times10^{-15}$erg$\,$s$^{-1}$cm$^{-2}$.
[SII]$\lambda\lambda 6716,6731$/H$\alpha$ shows a similar trend. }
\end{figure}

\section{DISCUSSION}

As shown in Fig. 2, the X-ray spectrum of \AX is unusually hard
compared with typical Seyfert galaxies and quasars, and similar in
shape to that of the XRB.  The current ASCA data is insufficient to
distinguish whether the X-ray spectrum is intrinsically flat or
absorbed (see Table 1), but the lack of significant X-ray detection in
the ROSAT soft band ($<0.4$ keV) suggests that an absorbed
Seyfert-type source is a likely interpretation.

The spatial variations in the line ratios also suggest the presence of
an active nucleus.  Similar trends in line ratios have been noted in
the starburst galaxy M82 (McCarthy et al. 1987), where the outer
regions of the nebula show LINER-like behaviour (although with
somewhat lower ionization than the peak in KUG\thinspace
1750+683A). This was attributed to shock heating in outflowing winds.
The X-ray emission from KUG\thinspace 1750+683A is unlikely to be due
to starburst activity however; to produce an X-ray luminosity of
10$^{43}$erg s$^{-1}$ would require starforming activity on an
unprecedented scale. This should then lead to a massive far infra-red
flux from the thermal re-radiation of dust heated by young stars.
Griffiths \& Padovani (1990) found a strong correlation between the
X-ray and 60$\mu$m luminosities in a large sample of star forming
galaxies. Extrapolating this relationship to KUG\thinspace 1750+683
would imply a 60$\mu$m flux density of $\sim30$Jy.  Assuming the same
X-ray to infra-red ratio as M82 would require a 60$\mu$m flux density
of $\sim20$Jy for KUG\thinspace 1750+683A.  Both values are grossly
inconsistent with the non-detection of this source by IRAS, with an
upper limit of only $\sim 0.1$Jy.  Similar conclusions are reached
when we compare with more recent ASCA observations of starburst
galaxies (Kii et al. 1996).  Thus the X-ray emission from AXJ1749+684
cannot be due to starburst activity.

If this is an obscured AGN, with an intrinsic photon index typical of
QSOs ($\Gamma=2$), then the X-ray spectrum implies an obscuring column
of $N_{\rm H}=8\pm3\times 10^{21}$\psqcm. Assuming standard
gas-to-dust ratios, this corresponds to a visual extinction of $A_V
\sim 5$ (Bohlin, Savage \& Drake 1978). In order to determine whether
this could extinguish the broad line region, we use the relationship
found by Ward et al. (1988) between the intensity in broad H$\alpha$
and the 2--10~keV X-ray luminosity. This predicts an intrinsic broad
H$\alpha$ flux of $\sim1\times 10^{-13}$ \ergpspsqcm. Once attenuated
by 5 magnitudes of extinction it is clear that such a broad line
component would be undetectable in the observed optical spectrum. It
is worth noting that the narrow-line Balmer decrement implies
$A_V\sim3$, and hence only a further $\sim 2$ magnitudes of additional
extinction are required to obliterate the broad-line region.

A similar hard X-ray spectrum was found in the spiral galaxy NGC3628,
which also shows a starburst/LINER like optical spectrum (Yaqoob et al
1995) but with a significantly lower X-ray luminosity than
AXJ1749+684.  Whatever causes these hard X-ray spectra, the detection
of weak X-ray sources like \AX supports the hypothesis that the
emission line galaxies emerging in deep ROSAT surveys could also
provide a solution to the origin of the hard XRB.  A puzzling feature
of NLXGs is that they tend to populate the borderline between
starburst and Seyfert 2 galaxies on the usual line ratio diagrams
(Ward et al 1993, Boyle et al 1995, McHardy et al 1997).  This
ambiguity is explained if these are moderately obscured AGN embedded
within active starforming galaxies, as seems to be the case for
KUG\thinspace 1750+683A.  Although we require the presence of an
obscured AGN to explain the large hard X-ray luminosity, as explained above,
the extended narrow emission-line spectrum and the detection of this
object in the Kiso Schmidt survey of UV excess galaxies must be due to
a large scale starburst.  Most of the NLXGs detected to date lie at
intermediate redshifts ($0.1<$z$<0.6$) where the field galaxy
population is well known to be undergoing significant starforming
activity (eg. Tresse et al 1996). The narrow optical emission lines
from an obscured AGN are then easily overwhelmed.  Spatially-resolved
optical spectra of NLXGs, such as presented here, should clarify the
presence of any emission-line region photo-ionized by an active
nucleus.

\section{CONCLUSIONS}

We have serendipitously detected a hard-spectrum X-ray source, \AXc,
which is identified with a spiral galaxy at a redshift of 0.05. This
galaxy shows an integrated LINER/starburst-like optical spectrum but
the X-ray luminosity is two orders of magnitude higher than any known
starburst galaxy.  Spatial variations in the optical emission lines
suggest the presence of a more ionized component.  In addition, the
lack of significant far infra-red emission is inconsistent with a
starburst origin for the huge X-ray flux. A plausible explanation is
the presence of a moderately obscured active nucleus surrounded by a
starforming galaxy. This would simultaneously explain the X-ray
emission and the optical spectrum.

The flat X-ray spectrum below 5~keV, combined with the optical
spectrum, suggest that this object may be a local counterpart to the
faint NLXGs thought to be responsible for the origin of the hard XRB.
This raises the intriguing possibility that many of these objects
could also contain AGN, embedded within starforming host galaxies. The
use of integrated optical spectra would then lead to an ambiguous
classification.  Many of the NLXGs currently classified as starburst
galaxies could well be more distant counterparts to \AXc.  Hidden AGN,
of which \AX is a low redshift and low column density example, may
therefore provide the origin of the hard XRB, as originally suggested
by Setti \& Woltjer (1989), and later modelled by Madau et al (1994)
and Comastri et al (1995).

\section*{ACKNOWLEDGEMENTS}

We thank all the members of the ASCA team who maintain the satellite
and carry out operations. This research has made use of the NASA/IPAC
Extragalactic Database (NED) which is operated by the Jet Propulsion
Laboratory, California Institute of Technology, under contract with
the National Aeronautics and Space Administration. The Digitized Sky
Surveys were produced at the Space Telescope Science Institute under
U.S. Government grant NAG W-2166. We thank R. Kollgaard and CDS at the
University of Strasbourg for providing the catalogue of the VLA NEP
survey, M. Tashiro and Y. Ishisaki for information about the GIS
positional accuracy and S. Ettori for help with the optical
observations.  ACF, CSC thank the Royal Society, WNB thanks the
Smithsonian Institution, and KI, OA thank the PPARC for support.

\bsp

\end{document}